\documentclass[twoside,openright]{nictatechreport}

\usepackage[T1]{fontenc}
\usepackage{flushend}

\newcommand{\labwiki}{{\sc LabWiki}\xspace}

\newcommand{\theTitle}{Repeatable Experiments with {LabWiki}}

\newcommand{\theKeywords}{reproducible experiments, testbeds, \labwiki, OMF, OML}

\usepackage{xspace}
\usepackage{csquotes}
\usepackage[australian]{babel}
\usepackage[maxnames=4,style=numeric,firstinits=true,sortcites,backend=bibtex,url=false]{biblatex}  

\DeclareNameAlias{default}{last-first}

\AtBeginBibliography{%
}

\DeclareFieldFormat
  [article,inbook,incollection,inproceedings,patent,thesis,unpublished]
  {title}{#1}

\renewbibmacro{in:}{%
  \ifentrytype{article}{%
  }{%
    \printtext{\bibstring{in}\intitlepunct}%
  }%
}

\renewbibmacro*{volume+number+eid}{%
  \printfield{volume}%
  \setunit*{\addcomma\space}%
  \printfield{number}%
  \setunit{\addcomma\space}%
  \printfield{eid}}

\DeclareFieldFormat{pages}{#1}

\renewbibmacro*{publisher+location+date}{%
  \printlist{publisher}%
  \setunit*{\addcomma\space}%
  \printlist{location}%
  \setunit*{\addcomma\space}%
  \usebibmacro{date}%
  \newunit}

\usepackage{graphicx} 
\usepackage{paralist}
\usepackage{amsmath}
\usepackage{listings}
\lstdefinestyle{narrow}{
  breakatwhitespace,breaklines,prebreak={\raisebox{0ex}[0ex][0ex]{\ensuremath{\hookleftarrow}}},
  keepspaces=true
  numbersep=5pt,
}
\lstdefinelanguage{sh}{basicstyle=\ttfamily}
\lstdefinelanguage{OEDL}{
  language=Ruby,
  escapeinside={\#latex:}{\^^M},
  emphstyle=\bf\sffamily,
  emph={loadOEDL,defProperty,defGroup,addApplication,setProperty,measure,
    onEvent,ALL_UP_AND_INSTALLED,group,startApplications,
  after,allGroups,stopApplications, Experiment,done},
}

\DeclareUrlCommand\fspath{\urlstyle{tt}}
\newcommand{\descr}[1]{\smallskip\noindent\textbf{#1}}
\usepackage{color}

\newcommand{\latinlocution}[1]{#1}
\newcommand{\eg}{\latinlocution{e.g.}}

\usepackage{epsfig}
\newcommand{\syscmd}[1]{\texttt{#1}}

\title{\theTitle}
\author{Thierry Rakotoarivelo,\textsuperscript{1,$\star$} Guillaume Jourjon,\textsuperscript1
Olivier Mehani\textsuperscript1\\
Max Ott,\textsuperscript1 Michael Zink,\textsuperscript2} 

\affiliation{$^\star$Corresponding author: \email{thierry.rakotoarivelo@nicta.com.au}\\
\textsuperscript1Nicta, Sydney. Eveleigh, NSW, Australia, \email{first.last@nicta.com.au} \\
\textsuperscript2Electrical and Computer Engineering Department, University of Massachusetts, Amherst, MA 01003, USA, \email{zink@ecs.umass.edu}}

\reportnumber{8350}
\copyrightyear{2014}
\headertitle{Rakotoarivelo {\em et al.} Repeatable Experiments with {LabWiki}}

\begin{document}

\begin{abstract}
The ability to repeat the experiments from a research study and obtain similar
results is a corner stone in experiment-based scientific discovery. This
essential feature has been often ignored by the distributed computing and
networking community.  There are many reasons for that, such as  the complexity
of provisioning, configuring, and orchestrating the resources used by
experiments, their multiple external dependencies, and the difficulty to
seamlessly record these dependencies. This paper describes a methodology based
on well-established principles to plan, prepare and execute experiments. We
propose and describe a family of tools, the \labwiki workspace, to support an
experimenter's workflow based on that methodology.  This proposed workspace
provides services and mechanisms for each step of an experiment-based study,
while automatically capturing the necessary information to allow
others to repeat, inspect, validate and modify prior experiments. Our \labwiki
workspace builds on existing contributions, and de-facto protocol and model
standards, which emerged from recent experimental facility initiatives. We use
a real experiment as a thread to guide and illustrate the discussion throughout this paper.
\end{abstract}


\frontmatter

\mainmatter
\section{Introduction}
\label{sec:intro}

One of the corner stones of scientific discovery is validation by the community.
In experimental science, it is the ability of others to repeat the experiments
and obtain similar results within acceptable statistical bounds. Traditionally,
the distributed computing and networking community has been largely ignoring
this. There are few publications in top-tier venues, which primarily report on
the successful validation of somebody else's work, while problems with
repeatability are sometimes buried in vague references. There are many reasons
for that. Advances in the underlying technology continuously create new
opportunities to explore new ideas leaving little time to reflect on the
``old''. But there are also very pragmatic reasons. First of all, most
experiments are conducted in complex environments with many external
dependencies, such as type and speed of computers and networks, size of storage,
chip sets, or operating system and driver versions. Some of them will only
affect the measured ``utility'' of the reported phenomena, while others are
essential to having a successful experiment in the first place. Unfortunately,
many of these dependencies are never reported and therefore making it very
difficult for others to repeat an experiment.

\begin{figure}[h]
\centering
 \includegraphics[width=0.40\textwidth]{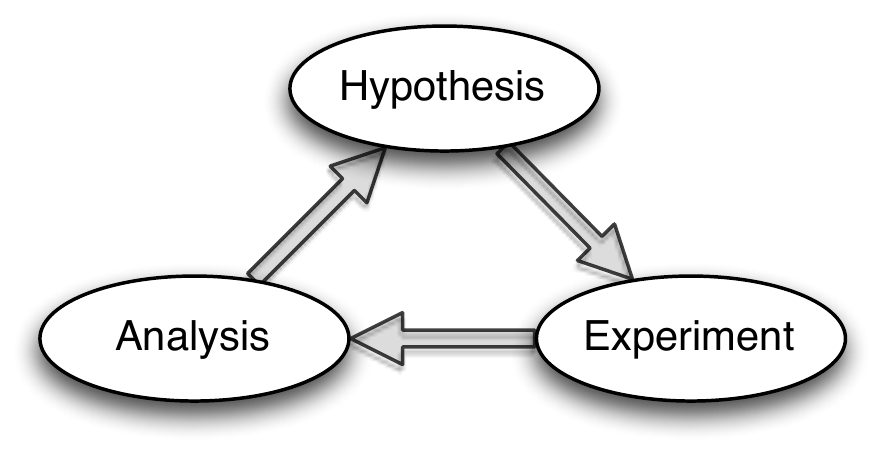}
\caption{Scientific Method.}
\label{fig:scientificMethod}
\end{figure}
We argue in this paper, that our inability to repeat reported experiments is not
only bad practice, but also hampers progress in general. It reduces our ability
to expand on prior work, verify and adapt it to different contexts, compare
different methods in different environments and much more. We also argue that
the ``paper'' as the traditional publication mechanism is one of the major
obstacles in improving the status quo.

We are clearly not alone, initiatives, such as the Elsevier's Executable Paper
Challenge~\cite{Gabriel2011} have been exploring new avenues for disseminating
scientific results. In addition, easy access to emerging large scale
experimental facilities, funded and coordinated by programs, such as GENI in
the US~\cite{berman2014}, FIRE in Europe~\cite{fire}, and similar activities in
China, Korea, and Japan, provide the community with a common ``playground'' in
which to conduct experiments. But that alone is not enough. The sharable
resources we have available now, still need to get provisioned, configured and
modified before they can be used in experiments. It is those steps that are
crucial to being able to repeat an experiment.

In the remainder of this paper we propose
and describe a family of tools to support an experimenter's workflow, while also
automatically capturing most of the necessary information to allow others to
repeat, inspect, validate and modify prior experiments.

\begin{figure}[t]
\centering
 \includegraphics[width=0.35\textwidth]{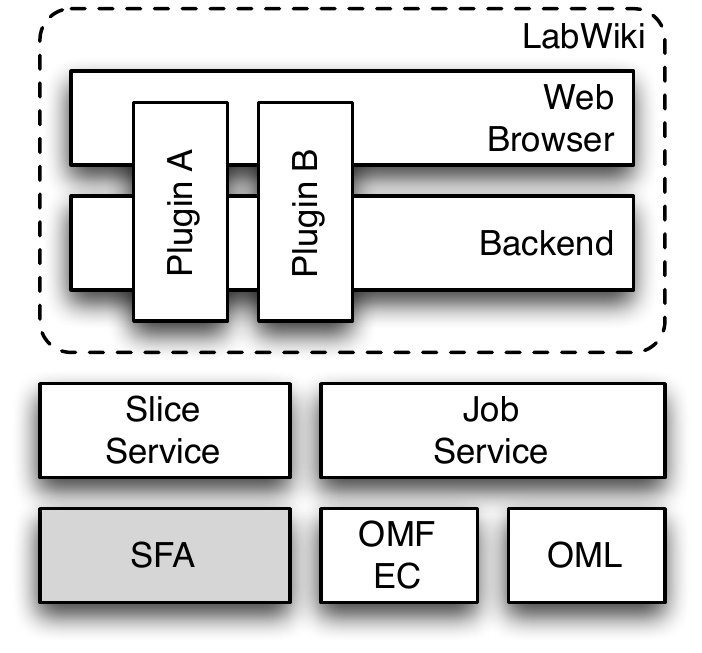}
\caption{\labwiki and Supporting Services.}
\label{fig:labwikiBlock}
\end{figure}

More specifically, we propose to model the experimenter workflow on the
Scientific
Method\footnote{\url{https://en.wikipedia.org/wiki/Scientific_method}} which we
interpret, as shown in Figure~\ref{fig:scientificMethod}, as a repeated cycle of
stating a hypothesis, designing and conducting an experiment, and finally
analyzing the measurements taken during the experiments with the intent to
disprove the hypothesis.  

We observed that many of these steps follow the same internal workflow of
planning, preparing, and executing. We therefore built an experimenter-facing
web-based tool, called \labwiki, which supports this three-step workflow in
different contexts.
\labwiki, as the name implies, is
modeled after the traditional laboratory book, which experimenters used for a
very similar workflow and purpose. \labwiki takes this further, by not only being
the recording mechanism, but also the operating platform for many activities
within the experiment workflow. We will 
 illustrate 
this process by taking the reader through every step,
namely the experiment
design (Section~\ref{sec:exp:design}), setup (Sections~\ref{sec:exp:desc:instr} 
and~\ref{sec:res:provision}), run (Section~\ref{sec:exp:run}) and 
analysis (Section~\ref{sec:result:analysis}) of a previously published 
research result~\cite{2014mehani_time_calibration_networked_sensors}. 

\labwiki, as shown in Figure~\ref{fig:labwikiBlock}, is sitting on top of a suite
of supporting tools and services, which can be used directly by an experimenter,
or more likely by other tools acting on her behalf. Specifically the
SliceService (Section~\ref{sec:sliceService}) which harmonizes resource
provisioning across many different testbeds; OEDL
(Section~\ref{sec:exp:desc:instr}), a domain-specific language  for describing
the orchestration of an experiment; JobService (Section~\ref{sec:exp:run}) for
scheduling an experiment; OMF \& FRCP (Section~\ref{sec:exp:run}) for executing
and coordinating individual experiment runs (or trials); and OML
(Section~\ref{sec:exp:desc:instr}) for collecting and managing measurements 
during a trial. We also briefly describe how the \labwiki workspace
can support educators in harnessing these large facilities for lab tutorials
(Section~\ref{sec:pub:arch}).



\section{LabWiki User Experience}
\label{sec:lw_ux}

\begin{figure}[t]
\centering
 \includegraphics[width=0.67\textwidth]{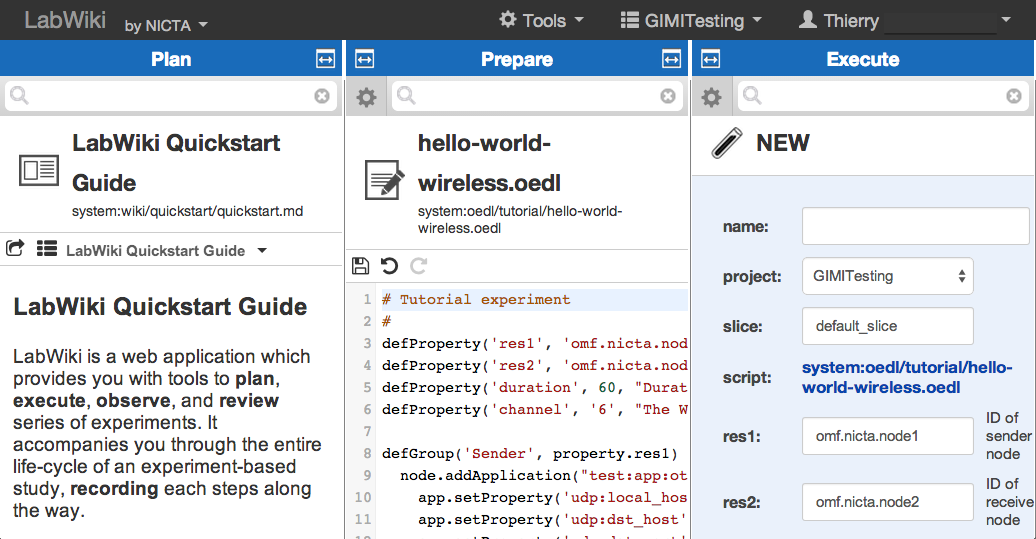}
\caption{Main interface of \labwiki.}
\label{fig:lw_main}
\end{figure}

As mentioned in the Introduction, the experimenter interacts with \labwiki primarily through a web browser. After a standard login process, the user will see (Figure~\ref{fig:lw_main}) a browser window split into three columns, labeled ``Plan'', ``Prepare'', and ``Execute''. This reflects the basic workflow identified above. Each column comprises of a tool \& search bar, followed by a widget header, an optional widget toolbar and the widget body. The top tool \& search bar allows the user to quickly locate or create resources relevant to the respective activity and choose the desired widget to interact with that resource. \labwiki itself is a framework with most of the functionality provided by plugins, which in turn provide one or more widgets. For instance, the \emph{wiki} widget for the ``Plan'' column supports editing of rich text resources.

All widgets are stateless and only provide mechanisms for a user to interact with one or more named resources. These resources may reside on a separate service, such as the JobService (Section~\ref{sec:JobService}), or are file-like resources, such as a wiki entry, or an image. For these kinds of objects, \labwiki provides layered, pluggable artifact stores. Current implementations support persistence through the local filesystem, versioned and access-controlled repositories such as
Git~\cite{2009chacon_pro_git}, and via iRODS~\cite{2010rajasekar_irods_primer}. The clean separation of stateless widgets and state-full, externally resolvable resources allows for interacting and embedding of these resources outside of \labwiki as well. For instance, plots of experiment measurements, hosted on JobService (Section~\ref{sec:JobService}) can be embedded into a wiki page, which then can be published from the  \emph{wiki} widget to a third-party blog service. Importantly, the link from the plot in the blog entry to the actual experiment is maintained, including access control mechanisms.

\labwiki supports multiple user accounts and uses OpenID for authentication. Resources, managed through \labwiki belong to projects and a user's membership and role in a project are the basis of {\labwiki}'s authorization mechanism. Information about membership and respective roles are sourced from external services, such as the GENI ClearingHouse. Currently \labwiki is also facilitating the transfer of delegation and \emph{speaks-for} credentials for the services some of the plug-ins call upon (e.g. SliceService~Section ~\ref{sec:sliceService}). 

\section{Experiment overview}
\label{sec:exp:overview}


As the main objective of \labwiki is to support a group of researchers in
producing verifiable experiments, we will use a real research experiment as the
guide through the reminder of this paper. This experiment was first designed
as part of a research effort on time synchronization in networked  sensors, with
the results published in~\cite{2014mehani_time_calibration_networked_sensors}.


Researchers in many domains, such as human-computer interaction, are
increasingly collecting large amounts of data from heterogeneous distributed
sensors. Accurately synchronizing these data streams is crucial for  meaningful
analysis and conclusions. While there are many, well-established  techniques for
synchronizing clocks in distributed
entities~\cite{2005roemer_time_synchronisation_calibration_WSN,rfc5905}, they
require additional software to be deployed on these entities, or depend on variables
which may not be under the experimenter's control (e.g., the offset between a NTP
client and a server depends on the network's round-trip delay). The above
mentioned research project proposed a different approach based on measurements
of the data collection system itself and uses the obtained meta data  to
synchronize the original data a-posteriori.



\begin{figure}[t]
\centering
 \includegraphics[width=0.67\textwidth]{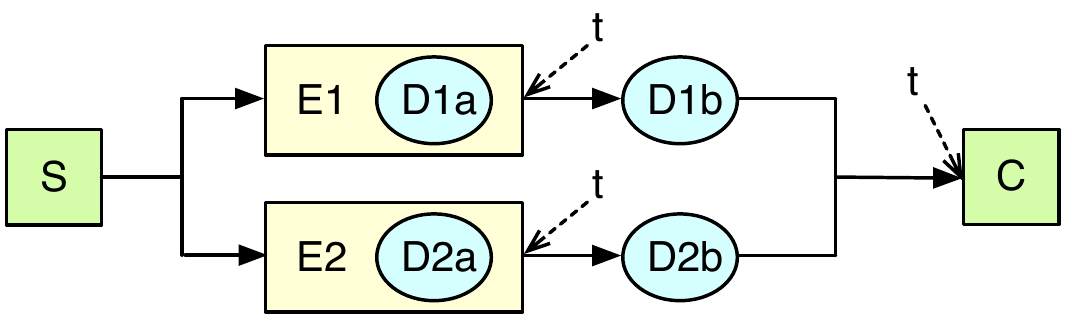}
\caption{Topology of the Time Calibration experiment from~\cite{2014mehani_time_calibration_networked_sensors}.}
\label{fig:exp:scenario}
\end{figure}

The main experiment assumes a scenario where certain events can be measured by
more than one sensor and where all sensors  then forward these measurements to a
common collection server.  Figure~\ref{fig:exp:scenario} illustrates the
resulting experiment topology. A series of events are  generated by a source $S$
and measured by two entities $E1$ and $E2$. The respective measurement samples
are sent to the same collection server $C$. Time delays may  be added at the
various $Dij$ points. $E1$, $E2$, and $C$ add locally sourced timestamps $t$ to
all samples that they produce and receive, respectively.

\section{Background}
\label{sec:relatedwork}

Major initiatives such as GENI~\cite{berman2014} and  FIRE~\cite{fire}
have focused on providing distributed, virtual laboratories for transformative,
at-scale experiments in network science and services. Designed in response to
the Internet ossification issue, these so-called \emph{testbeds} enable a wide
variety of experiments in many areas, including clean-slate networking, protocol
design, distributed service offerings, social network
integration, content management, and in-network service deployment. Many
software tools were proposed to allow operators and experimenters to manage,
access and control the resources from these testbeds.  The models, protocols,
and APIs from some of these contributions are currently converging towards 
\emph{de-facto} standards within the community.

%

A RSpec\footnote{\url{http://geni.net/resources/rspec}} defines a set of
resources that can be used in an  experiments. These resources may be
requested from a variety of GENI or FIRE testbeds, such as ExoGENI, OpenFlow
Mesoscale, or
Fed4FIRE~\cite{2013vandenberghe_architecture_heterogeneous_federation_testbeds}.
There are three different types of RSpecs, \begin{inparaenum}[\itshape
a\upshape)] \item \emph{the Advertisement} which is sent by an Aggregate Manager
(AM) to an experimenter to describe its available resources; \item \emph{the
  Request} which is sent by  the experimenter to the AM to describe the
resources she wants to reserve; and \item \emph{the Manifest} which  is returned
  by the AM to describe which resources have been reserved by the experimenter.
\end{inparaenum} These RSpecs are exchanged in the previous sequence  between the
AM and the experimenter. The requested resources will be available to  the
experimenter after the successful completion of that sequence.


The Aggregate Manager APIs~\cite{berman2014} define a common interface for
software to provide, request, reserve, and provision resources over different
facilities. They are based on a \emph{slice} abstraction, which is a container
for all the resources used in a project. Experimenters are associated with
slices and use these APIs to interact with various entities (e.g. Clearinghouse,
Aggregate Manager) in order to discover, reserve and provision resources. These
interactions are mostly performed through third-party interfaces. For example,
Omni\footnote{\url{http://trac.gpolab.bbn.com/gcf/wiki/Omni}} is a command line
tool used to specify and reserve resources from GENI facilities. It allows
\emph{stitching}, a technique to connect resources via layer 2 VLans. In
contrast, Flack and Jack\footnote{\url{http://www.protogeni.net/wiki/Flack}} are
graphical tools, which allow experimenters to reserve resources and specify
RSpecs through a visual topology editor. Finally,
JFed\footnote{\url{http://jfed.iminds.be}} is a Java-based tool, which allows
experimenters to obtain large distributed topologies using resources from both
FIRE and GENI testbeds.

The Federated Resource Control Protocol (FRCP) and the OML Measurement Stream
Protocol\footnote{\url{http://oml.mytestbed.net/doc/doxygen/omsp.html}}
(OMSP)~\cite{2012rakotoarivelo_omf,2014mehani_instrumentation_framework} are two
protocols to control resources and collect data from them. They are commonly
used in both GENI and FIRE facilities. FRCP defines a short set of asynchronous
interactions over a \emph{publish-and-subscribe} system, which allows
experimenters to configure resources and instruct them to execute given tasks.
The OMF and NEPI control tools both implement
FRCP~\cite{rakotoarivelo2010,nepi2014}. OMSP defines the format and transport of
measurement tuples from producers (\eg, a resource) to consumers (\eg, a storage
server). It supports various types of measurements, encodings, and the use of
metadata. The OML framework~\cite{2014mehani_instrumentation_framework} provides
an OMSP storage server and a C client library to instrument resources. Other
client libraries also exist (\eg,
OML4R,\footnote{\url{https://github.com/mytestbed/oml4r}}
OML4Py\footnote{\url{https://github.com/mytestbed/oml4py}} or
OML4J\footnote{\url{https://github.com/NitLab/oml4j}}).

\section{Experiment design}
\label{sec:exp:design}

The first step of an experimental study is the design of the experiment itself. It is driven by 
 research goals, such as testing a hypothesis, measuring performance, or
demonstrating capabilities.


There have been many contributions related to experiment design  since the
seminal work of Fisher~\cite{fisher1937}. Examples in the area of computer
science include~\cite{paxson2004,krishnamurthy2011}. While there are many
variations, a good starting point is the identification of the dependent, independent,
and confounding variable sets. The \emph{dependent} variables
are measured attributes of the studied system, their analysis will provide answers to the study's questions.
The \emph{independent} variables
would impact the studied system and modify its dependent variables. The third
set of \emph{confounding} variables  may be unknown or uncontrollable by the
experimenter and  may have some effect on any of the former variables.

Given these three variable sets, the researcher then devises an experiment plan where
usually the \emph{dependent} variables are measured, the \emph{independent}
variables are controlled and varied across different repeated trial batches,
and the effect of \emph{confounding} variables are mitigated through techniques
such as replication or randomization. The choice of controlled values for the
\emph{independent} variables and the number of trials and their repetition
depends on the objectives of the study.

The \labwiki workspace has a set of tools to support the experiment
design process. The ``Plan'' column on the left-hand side of its interface
(Figure~\ref{fig:lw_main}) provides a Wiki widget  that allows the 
 experimenter  to describe and record her design. This design
strives to replace her pen-and-paper laboratory notebook. It currently uses the popular
Markdown syntax\footnote{\url{http://daringfireball.net/projects/markdown}},
and figures and plots from other widgets can be easily dragged-and-dropped into the
write-up.

\descr{The Design of Our Example Experiment.} 
In the case of our example experiment, we identify the \emph{dependent} variables
as the arrival times of a measurement sample at different points in the
system.
Our \emph{independent}
variables consist of configurable clock offsets and network delays, generally
referred to as $Dij$ in Figure~\ref{fig:exp:scenario}. One
potential \emph{confounding} variable would be the varying delays in processing
measurement samples inside the sensors.

In this particular realization of the experiment,
we chose a classic ICMP ping to the network's broadcast address as the
\emph{event} generated by the source $S$ and measured through their
respective network
interfaces on $E1$ and $E2$.
In the first round of trials, the entities are quasi-synchronized 
and no delay is added on the collection network. This establishes a
baseline for future results. We then planned to run
a series of trials, where various known time offsets are introduced to
each entity's clock and on their respective paths on the collection network. To
mitigate the potential impacts of \emph{confounding} factors, we decide to run
multiple trials for each specific offset configuration and further repeat
these multiple trials over different instantiations of our topology. 
More details about this experiment design and each series of trials are
available in the original
study~\cite{2014mehani_time_calibration_networked_sensors}.

As mentioned above, we primarily use the Wiki widget to describe the design and 
work plan. Dragging experiment results and other artifact onto the wiki will allow us 
to keep track of progress. This will be especially important if an experiment is carried
out by a team where different members are pursuing different parts of the work plan.



\section{Experiment description and instrumentation}
\label{sec:exp:desc:instr}


Once the design of the experiment is finalized and documented in {\labwiki}'s Plan 
panel, the next step is to translate it into a machine-readable description.

\subsection{Describing an Experiment}
\label{sec:exp:desc:instr:desc}

We propose to use the existing OEDL language~\cite{rakotoarivelo2010} to describe an experiment.
OEDL has been widely used to describe repeatable experiments on both GENI and
FIRE facilities. A typical OEDL script is primarily composed of two sections. The first
one declares the resources used in the experiment and their initial
configurations. For example, an experimenter may declare a given application to
be used, along with its available parameters and
measurement capabilities; and the specific initial settings for both of them.
The second section of the OEDL script defines the orchestration of tasks 
the resources have
to execute throughout an experiment trial. These tasks are grouped into experimenter-
defined events, which may be triggered either by timers or experiment-specific
conditions. This event-based approach allows complex experiment orchestrations,
such as changing the parameters of an application X seconds after the measurement of
Y from another application reaches the value Z.

\labwiki has inspired a third, optional section for a typical OEDL script. It allows
an experimenter to define charts to provide quick feedback on the progress 
or outcome of an experiment trial.
The \emph{experiment} widget in \labwiki uses that to display line, pie,
or histogram charts
in the respective column with relevant measurement data streams sourced from the JobService.
We do want to note, that this is primarily 
to provide a graphical live feedback on an
individual execution of an experiment trial, and will usually not replace a
thorough data analysis over the complete result set obtained for an experiment
(Section~\ref{sec:result:analysis}).

\lstlistingname~\ref{lst:test-sync} provides a shortened OEDL script for our
example experiment. While the complete OEDL
script\footnote{\url{http://git.io/clock-delay-impairments.rb}} describes the
entire experiment with all required
settings as designed in Section~\ref{sec:exp:design}, this shortened version only shows the experiment for the baseline trials.
We will now briefly describe 
this script and refer the reader to the
OEDL Reference document\footnote{\url{http://omf6.mytestbed.net/OEDLOMF6}} for
further details. 

\begin{lstlisting}[language=OEDL,
  caption=Example of an OEDL script,
  label=lst:test-sync,
  float=tbh,
  numbersep=5pt,
  ]
loadOEDL('http://goo.gl/4br2MW') #latex:\label{lst:test-sync:load-begin}
loadOEDL('http://goo.gl/qg8Alo')
# From the trace-oml2 package
loadOEDL('file:///usr/share/trace-oml2/trace.rb')#latex:\label{lst:test-sync:load-end}

defProperty('entity1','node20','1st entity ID') #latex:\label{lst:test-sync:prop-begin}
defProperty('entity2','node21','2nd entity ID')
defProperty('source','node19','Event source ID')
defProperty('time',180*60,'Trial duration [s]') #latex:\label{lst:test-sync:prop-end}

defGroup('Entities',prop.entity1,prop.entity2) do |g| #latex:\label{lst:test-sync:entities-begin}
  # Capture ICMP echo packets
  g.addApplication('trace') do |app| #latex:\label{lst:test-sync:trace}
    app.setProperty('filter', 'icmp[icmptype]=icmp-echo')
    app.setProperty('interface', 'eth1')
    app.measure('ethernet', :samples => 1) #latex:\label{lst:test-sync:measure-ethernet}
  end
end #latex:\label{lst:test-sync:entities-end}

defGroup('Source',prop.source) do |g| #latex:\label{lst:test-sync:source-begin}
  # Broadcast ICMP echo requests every 10s
  g.addApplication('ping') do |app| #latex:\label{lst:test-sync:ping}
    app.setProperty('dest_addr', '10.0.0.255') #latex:\label{lst:test-sync:pingbroadcast}
    app.setProperty('broadcast', true)
    app.setProperty('interval', 10)
    app.setProperty('quiet', true)
    app.measure('ping', :samples => 1) #latex:\label{lst:test-sync:measure-ping}
  end
end #latex:\label{lst:test-sync:source-end}

defGroup('All',prop.source,  #latex:\label{lst:test-sync:all-begin}
          prop.entity1,prop.entity2) do |g| 
  # Report time synchronisation every minute
  g.addApplication('ntpq') do |app| #latex:\label{lst:test-sync:ntpq}
    app.setProperty('loop-interval', 60)
    app.setProperty('quiet', true)
    app.measure('ntpq', :samples => 1) #latex:\label{lst:test-sync:measure-ntpq}
  end
end #latex:\label{lst:test-sync:all-end}

onEvent(:ALL_UP_AND_INSTALLED) do #latex:\label{lst:test-sync:experiment-begin} 
  group('All').startApplications
  group('Entities').startApplications
  group('Source').startApplications
  after prop.time do
    allGroups.stopApplications
    Experiment.done
  end
end #latex:\label{lst:test-sync:experiment-end}

\end{lstlisting}

Lines~\ref{lst:test-sync:load-begin}--\ref{lst:test-sync:load-end} fetch and
load additional OEDL scripts, similar to the \emph{include} statement found in
many programming languages.
Lines~\ref{lst:test-sync:prop-begin}--\ref{lst:test-sync:prop-end} define some
experiment parameters which may be modified for different trials.
Lines~\ref{lst:test-sync:entities-begin}--\ref{lst:test-sync:entities-end}
define a group of resources comprising of the entities $E1$ and $E2$ from
Figure~\ref{fig:exp:scenario}. An ICMP packet capture application is associated
to each resource in that group (l.~\ref{lst:test-sync:trace}). The
parameters and measurements to collect for this application are set using the
\lstinline[language=OEDL,basicstyle=\normalsize]{setProperty} and
\lstinline[language=OEDL,basicstyle=\normalsize]{measure} commands,
respectively.
Lines~\ref{lst:test-sync:source-begin}--\ref{lst:test-sync:source-end} define
another group of resource with only the source $S$ from
Figure~\ref{fig:exp:scenario}. The ICMP ping application is associated to that
resource (l.~\ref{lst:test-sync:ping}), and configured to ping the network's
broadcast address (l.~\ref{lst:test-sync:pingbroadcast}).
Lines~\ref{lst:test-sync:all-begin}--\ref{lst:test-sync:all-end} define a third
group which include all the previous resources. A time statistic reporting
application is associated to all these resources (l.~\ref{lst:test-sync:ntpq}).
Finally,
Lines~\ref{lst:test-sync:experiment-begin}--\ref{lst:test-sync:experiment-end}
define the sequence of tasks to perform once all the resources are ready and
their associated applications are installed. In this simple case, all the
resources first start their time reporting applications. Then the resources
within the ``Entities'' group start their applications. Then the resource in the
``Source'' group does the same. After a set duration, all resources in all the
groups stop their applications, and the trial is finished.

\subsection{Instrumenting Resources}
\label{sec:exp:desc:instr:oml}

\begin{figure}[bt] \resizebox{\columnwidth}{!}{\input{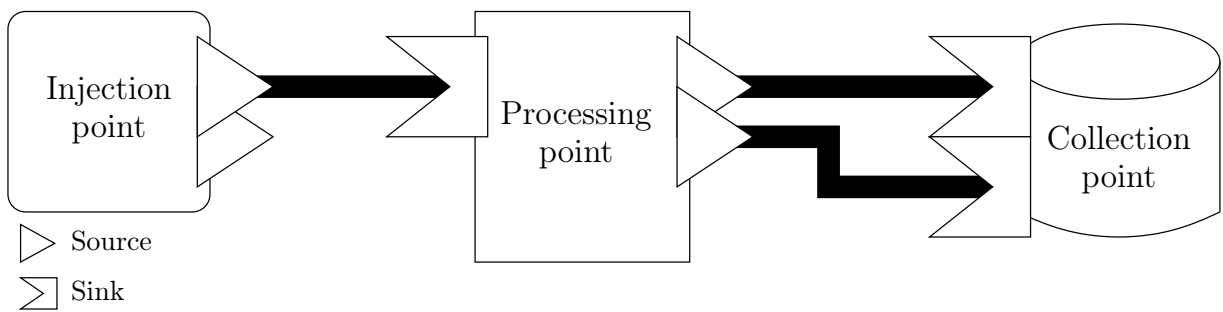}}
  \caption{Applications can be instrumented with OML to \emph{inject}
  timestamped samples into various measurements streams (MS) which can be
  \emph{processed} in-line (\eg, aggregated or sub-sampled) before finally being
  \emph{collected} and written into a storage backend.}

  \label{fig:omsp}
\end{figure}

\labwiki leverages the OML measurement
framework~\cite{2014mehani_instrumentation_framework} for 
measurement collection and storage.  OML is based on the concept of
\emph{measurement points} (MPs). The schema of an MP defines a tuple of typed                                                          
metrics meaningfully linked together 
(\eg, sampled at the same time, or pertaining to the same group). The series of
tuples generated by reporting measurements through an MP defines a
\emph{measurement stream} (MS). OML defines several entities along the
reporting chain that can generate, manipulate, or consume MSs. This is
illustrated in \figurename~\ref{fig:omsp}.

\subsubsection{Instrumentation Process}

The instrumentation of a resource consists of enabling applications to act as
injection points (\figurename~\ref{fig:omsp}).  By providing a structured way of
defining MPs, OML fosters the reusability and interchangeability of instrumented
applications, and simplifies the assembly of subsequent experiments. For example
the ``wget'' and ``cURL'' applications report similar information about web transfers,
and should therefore attempt to reuse the same MPs.

It is therefore important to first identify all the measurements that can be
extracted from an application. For example, \syscmd{ping} not only provides
latency information, but also sequence and TTL information for each received
packet from any identified host, as well as overall statistics. A rule-of-thumb
is that measurements, which are calculated, measured or printed at the same
time/line are good candidates to be grouped together into a single MP.  For more
complex cases, where samples from multiple MPs need to be linked together, OML
provides a specific data type for \emph{globally unique identifiers} (GUID).
They can be used in a similar way as foreign keys in databases.  For example, in
the case of the \syscmd{trace-oml2} application, it was decided to create one MP
per protocol encapsulated in a packet (\eg, ethernet or IP). A fresh GUID is
generated for each packet, which is then parsed, injecting information about each
header in the relevant MP, along with the GUID.

It is also possible to report metadata about the current conditions. Such
details as description, unit  and precision of the fields of an MP are
primordial for the later understanding of the collected
measurements.\footnote{Base SI units should be preferred whenever possible.}
Other information such as the command line invocation, or application version
and parameters are also worthy of inclusion as part of this metadata.

%

\subsubsection{Instrumentation Libraries}

\begin{sloppy}
  The most complete OML implementation is the C
  \syscmd{liboml2}(1,3).\footnote{Manpages for OML system components can be
    found at \url{http://oml.mytestbed.net/doc}.} It provides an
  API, which can be used to define MPs within the source code of an application,
  and mechanisms to process the injected samples at the source. The
  \syscmd{oml2-scaffold}(1)
  tool can be used to generate most of the boilerplate instrumentation code, along
  with the supporting OEDL
  description~\cite[App.~A]{2014mehani_instrumentation_framework}. An example
  application written from scratch to report network packets is
  \syscmd{trace-oml2},\footnote{\url{http://git.mytestbed.net/?p=oml-apps.git;a=blob;f=trace/trace.c}}
  used in \lstlistingname~\ref{lst:test-sync} (l.~\ref{lst:test-sync:trace}).

\end{sloppy}

The Ruby and Python bindings (OML4R \& OML4Py) are particularly useful for
writing wrappers for applications for which the source code is unavailable.
Wrappers work by parsing the standard output of an application, and extracting
the desired metrics to report. An example is the \syscmd{ping-oml2}
wrapper\footnote{We generally use \syscmd{APPNAME-oml2} as the binary's name of
OML-instrumented versions of upstream \syscmd{APPNAME} utilities; the OEDL
application description however only uses
\lstinline[basicstyle=\normalsize]{APPNAME} for conciseness.}, using
OML4R\footnote{\url{http://git.io/oml4r-ping-oml2}}, used in
\lstlistingname~\ref{lst:test-sync} (l.~\ref{lst:test-sync:ping}).


\subsection{The Prepare Panel}

Our \labwiki workspace has a ``Prepare'' panel at the center of its interface
(Figure~\ref{fig:lw_main}), which provides a simple code editor widget. The
experimenter may use this widget to edit an OEDL script, as
described previously. 
All OEDL scripts created within this editor widget
are versioned and saved within {\labwiki}'s artefact store with group-based
access control. While this widget is a convenient tool for users to edit their
OEDL scripts, the may use alternate means to do so, as well. For example, they
may edit their scripts in other  editors and then cut \& paste it into the
``Prepare'' panel's code editor 
Alternatively, they may 
directly use a git repository, and push it into {\labwiki}'s artefact store. 
\section{Resource selection and provisioning}
\label{sec:res:provision}

While the previous section dealt with describing the entire experiment and its
resource needs, it did not consider where these resources come from. 
For instance,  the
OEDL script in \lstlistingname~\ref{lst:test-sync}  refers to resources,  such
as virtual machines (nodes) and applications (trace, ping).  It is the former
kind of resources, which we assume will be provided by  testbeds and the
programmable networks connecting them.


More specifically, an experimenter needs to first define a topology of nodes,
their interconnecting networks, and their specific characteristics. For our
example  experiment, at least four (virtual) machines and four links are required
to create the topology in Figure~\ref{fig:exp:scenario}.
We note, that should the experiment be extended (e.g., by adding new entities E3
and E4), additional resources have to be reserved. Alternatively, the
experimenter may reserve a larger  topology and  run different trials on a
subset of the reserved resources.

\subsection{Process Overview}

\descr{Specification.}  The very first step is the specification of resources
that are required for an experiment. Section~\ref{sec:relatedwork} described several approaches 
for resource description
and tools to create them. 
The most common specification is the XML-based GENI RSpec\footnote{\url{http://groups.geni.net/geni/wiki/GENIExperimenter/RSpecs}}.

\descr{Reservation.} Once specified, a set of resources needs to be reserved.
This requires a negotiation phase between the provisioning tools and the corresponding services 
on the testbed side. 
A negative outcome of this negotiation means
that the requested resources can currently not be provisioned, or the requestor does not have 
sufficient privileges, or has exceeded her quota. For example, a VM
is requested but all the physical machines' resources have been already
allocated to other experimenters. In such a case, the experimenter either waits
or hopes that the desired resources will become available in the near future
or modify the request for a different set of
resources. A positive response means that the provisioning process will move on
to the next step.

\descr{Provisioning.} After specific resources have been identified, they need to be provisioned
before the experimenter can gain access to them. In the case of a physical machine, this may
require a power up. In the case of a VM, a disk image containing the requested operating system 
needs to be loaded and the VM started up with the appropriate configurations. This may also include
the distribution of security credentials to limit access to those resources to the requesting party.
 As each step may take time and in the case of large requests never fully complete successfully, proper communication between the requesting and providing services need to be maintained as even a subset of successfully provisioned services can already be used for successful experimentation.


%

\descr{Monitoring.} Most of the resources provided are virtualized in some form or depend on other services
in non-obvious ways. It is therefore important for most experiments to be able to monitor their resources and 
potentially even the broader context in which they are provided. For instance, CPU and memory allocations to 
VMs may change over time, or there may be external interference in wireless testbeds. While some of these
parameters can be monitored by the experimenter herself, others may need special access  and therefore need
to be collected by the resource provider with the results forwarded to the experimenter. For instance, the BonFIRE~\cite{bonfire} testbed provides monitoring information on the physical server to the VM ``owners'' for the respective server.
An experimenter can either use such infrastructure information after the
completion of the experiment during the analysis phase or display this
information in real time in LabWiki for actual monitoring.

%

\subsection{Labwiki Topology Editor Plugin}
\label{sec:lw_plugin}
	
\labwiki provides a
topology editor plugin, which supports the experimenter in navigating the above described steps. The plugin provides two widgets, one for the ``Prepare'' panel to specify the topology, and one for the ``Execute'' panel to request the provisioning of a defined topology and its ongoing monitoring.

\begin{figure}[t]
\centering
 \includegraphics[width=0.5\textwidth]{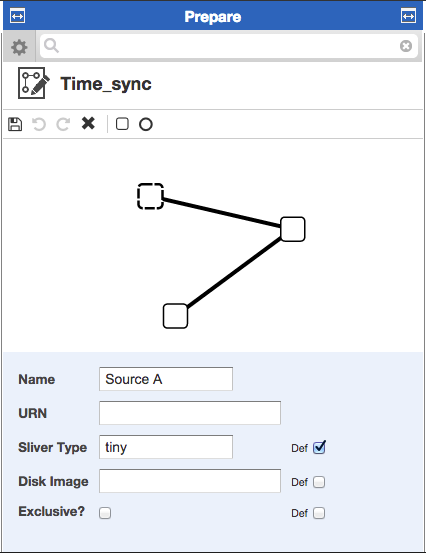}
\caption{\labwiki's topology widget.}
\label{fig:lw_topology_widget}
\end{figure}

Figure~\ref{fig:lw_topology_widget} shows a screenshot of the first widget, the
graphical topology editor. The widget is split vertically with the graph editor
on top and the property panel for the selected resource (dotted outline) at the
bottom. Interactive graphical editors are usually easy to learn, but do not
scale well to large topologies. Hierarchical grouping with associated visual
``collapsing'' can mitigate some of these scaling issues. However, larger
topologies will not be ``hand crafted'' but generated by tools, such as
BRITE~\cite{brite}. The topology editor has a text-mode, which allows the
experimenter to specify a BRITE model as well as provisioning information for
the nodes  and links created.

The topology description can either be stored as RSpec or extended GraphJSON. It is this textual representation, which the ``Slice'' widget is sending  to the SliceService when requesting the reservation and provisioning of a specific topology. This widget uses the same graph editor (now read-only) to convey progress by animating the graph elements accordingly. Monitoring information is also overlaid to provide experimenters with quick feedback on the overall topology status.

\subsection{The SliceService}
\label{sec:sliceService}

The SliceService provides a REST API for requesting and provisioning of resources for a testbed federation. It is essentially a proxy service to the SFA APIs of the ClearingHouse and AggregateManager. We reluctantly chose this path as the legacy decisions regarding technology (XML-RPC and client-authenticated SSL), as well as multiple versions for both API and RSpec put a considerable maintenance burden on the upstream tools. Therefore, we designed and implemented a service, which is based on current best practices for web-based services. We want to stress, that this is not a value judgement of the SFA APIs but a commonly encountered legacy problem. This decision allows us to concentrate our development as well as debugging efforts regarding testbed interactions on a single service. In fact, some design decisions for the SliceService have been heavily influenced by JFed\footnote{\url{http://jfed.iminds.be}}, which seems to have similar objectives.

Following the REST philosophy, SliceService defines a distinct set of resources and provides a consistent set of actions to create, modify, and delete those resources. It also takes advantage of the recently introduced delegation mechanism based on credentials. In fact, SliceService in combination with \labwiki seems to be one of the first services taking advantage of this capability. 

The SliceService also plays a crucial role in the security mechanism of FRCP (Section~\ref{sec:relatedwork}) by providing the newly created resources with proper credentials in a secure manner. However, a detailed description of the overall security mechanism is beyond the scope of this paper.

\section{Running an experiment trial}
\label{sec:exp:run}

Once an experiment is designed, described, and all necessary resources have been allocated,
the next step is to execute an instance (or trial) of that experiment.
Running an experiment trial should be effortless for the experimenter, as she
will need to repeat this process many times in order to gather sufficient data
to derive statistically meaningful conclusions.

\subsection{The Execute Panel}
\label{sec:lw:plan}

Our \labwiki workspace has an ``Execute'' panel on the right-hand side of its
interface (Figure~\ref{fig:lw_main}), which hosts an experiment widget. This
widget allows an experimenter to configure,
start, and observe individual experiment trials. To initiate this process, an experimenter
intuitively drags \& drops the experiment's icon from the previous ``Prepare''
panel into the ``Execute'' one. This action triggers \labwiki to display 
a list of the experiment properties defined in the respective OEDL script
which can now be configured for a specific trial. For example, the experiment design
might require that 20 trials should be ran with property A set to 1, followed by
another 20 trials with A set to 2. Once the experimenter is satisfied with
the trial's configuration, she may press the start button at the bottom of the
panel, which instructs {\labwiki}'s experiment widget to post a request for 
trial execution to an external JobService instance.

\subsection{The JobService and Its Scheduler}
\label{sec:JobService}

Our \labwiki workspace de-couples
the frontend interface used to develop and interact with experiment artifacts
from the backend processes orchestrating the execution of an experiment trial. The
JobService software is the backend entity in charge of supervising this
execution.
This decoupling enables our tools to cater for a wide range of usage scenarios,
such as use of an alternative user frontend, automated trial request (e.g. software can
request a given trial to be run at a periodic time), optimization of a shared
pool of resources among trial requests from the same project.

The JobService provides a REST API, which allows clients such as a \labwiki
instance to post trial requests (i.e. experiment OEDL scripts, property
configuration). Each request is passed to an internal scheduler, which queues it
and periodically decides which job to run next. This scheduler function is a
plugable module of the JobService, thus a third party deploying its own JobService
may define its own scheduling policies. In its simplest form, our
default Scheduler is just a plain FIFO queue. However, in an education context it
could be a more complex function, which could allow a lecturer to optimize the
use of a pool of resources (allocated as in Section~\ref{sec:res:provision})
between parallel experiment trials submitted by multiple groups of students. The
JobService's REST API also allows a client to query for the execution status
and other related information about its submitted trials. \labwiki uses this
feature and displays the returned information in its ``Execute'' panel once the
trial execution has started.

\subsection{Orchestrating Resources}

The JobService uses the existing OMF framework~\cite{rakotoarivelo2010},
which is available on many GENI facilities to orchestrate experiment trials.
More specifically, when the JobService's Scheduler selects a given trial
request for execution, it starts an OMF Experiment Controller process (EC). This
EC interacts with a Resource Controller (RC) running on each of the involved
resources, and have them enact the tasks required in the experiment description.
This interaction is done via the  FRCP protocol (Section~\ref{sec:relatedwork}). This
approach enables interoperability with other third-party ECs or RCs, the support
for disconnections of any durations between controllers, and trials with large
number of resources. While the trial is being executed, the JobService
constantly monitors the information from the output of the EC process and uses
it as part of the status provided back to \labwiki.
While our current JobService uses OMF for its underlying experiment trial
execution, its design also permits the use of other alternative frameworks, 
such as NEPI~\cite{nepi2014}.

\subsection{Collecting Measurements}

The applications instrumented in Section~\ref{sec:exp:desc:instr:oml} inject
measurement streams from measurement points as selected by the experiment
description (\eg, \lstlistingname~\ref{lst:test-sync}
ll.~\ref{lst:test-sync:measure-ethernet} or \ref{lst:test-sync:measure-ping}).
In \figurename~\ref{fig:omsp}, the reporting chain is terminated by a collection
point. The OML suite~\cite{2014mehani_instrumentation_framework} provides an
implementation of this element in the form of the
\syscmd{oml2-server}(1).
It accepts OMSP streams on a configurable TCP port, and
stores the measurement tuples into SQL database backends.

In our \labwiki workspace, the EC instructs the applications on the location of
the collection points to report their MSs to.
One database is created per experiment,
and a table is created for each MP (regardless of how many clients report into
this MP). The \syscmd{oml2-server} server currently supports SQLite3 and
PostgreSQL databases, and there are plans to extend this to semantic and NoSQL
databases. 

For very large deployments, the OML collection can be scaled by running multiple
\syscmd{oml2-server}s behind a TCP load balancer such as
HAProxy.\footnote{\url{http://www.haproxy.org/}} Instrumented applications carry
all the necessary state information in the initial connection for any server to
create the tables and store the reported tuples. A PostgreSQL cluster can then
be used as a backend to store data into a single centralised logical location
where analysis tools can access data both in real time and retroactively.


\section{Result analysis over multiple trials} 
\label{sec:result:analysis} 

As mentioned in Section~\ref{sec:exp:design}, it is often necessary 
to execute several trials of a
given experiments 
to gather sufficient data for a meaningful
analysis. Our proposed \labwiki workspace facilitates these replicated trials,
and provides two options for the experimenter to use the produced data.

First, the ``experiment'' widget provides 
an ``Export'' button 
 once the respective trial has
terminated. 
Depending on the \labwiki configuration, 
data can be exported
either as a self-contained database file (SQLite3 or
PostgreSQL dumps), as a compressed (e.g. ZIP) archive of comma separated
(CSV) formatted files, or pushed into an existing iRODS data store. 
{\labwiki}'s plugin-based design
allows other third-party export widgets to be provisioned as requested. The experimenter
may then download the produced database and import it into her preferred data
analysis software.

\begin{figure}[t]
\centering
 \includegraphics[width=0.5\textwidth]{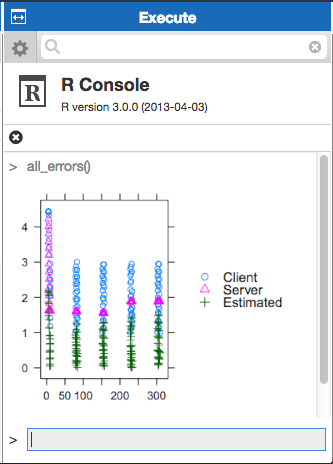}
\caption{\labwiki's analysis widget.}
\label{fig:analysis_widget}
\end{figure}

\labwiki provides another alternative to interact with the produced data through
another widget, which interfaces with a separate R server. This widget allows the
experimenter to nominate an R script to be submitted
to the R server, which executes the script's instructions and
returns any outputs (text or graphic results) to the widget for display.
Figure~\ref{fig:analysis_widget} shows a screenshot of this widget being used to analyze data from
our example experiments. 

In many cases, the result analysis of multiple trials will provide new insights
into the subject of the study. The experimenter may then reflect on these
conclusions and decide to either run further trials with an updated experiment design,
description, used resources, and/or analysis as required. This  step
effectively closes the experiment workflow process as described in
Section~\ref{sec:intro}.




\section{Store and Publish}
\label{sec:pub:arch}

As an experiment-based study progresses, it is essential to ensure that all generated artifacts, such as documents, data, and analysis, are permanently stored for continued and future access. Furthermore,
once a study finishes, the experimenter should be able to easily share her findings with the community.

\subsection{Storing and Sharing with \labwiki}
\label{sec:pub:storing}

As mentioned previously, \labwiki supports different types of artifact stores.
This allows different operators to deploy their own \labwiki instance, to select
the storage solution that best suits their requirements. For example, in an
evaluation or small closed institution context, a plain directory hierarchy
based on the underlying file system may be enough. In contrast, for a large
community spanning multiple organizations (e.g. GENI), a solution offering fine-
grain access control, high service availability, and version control may be
more appropriate. A list of the supported artifact stores 
are available in our source code
repository, together with \labwiki's ``Repository'' API, which could be used to add interfaces to other storage alternatives\footnote{\url{http://git.io/omfweb}}.

To faciliate the sharing of generated documents, \labwiki's ``wiki'' plugin has a pluggable
export component, which allows the publication of its documents to an external
Content Management System (CMS). This feature is accessed through an ``Export''
action displayed at the top of {\labwiki}'s ``Plan'' panel. There is currently only
one export plugin available, which allows the sharing of documents to a Respond
CMS system\footnote{\url{http://respondcms.com/}}. The source code for this
component along with documentation on how to implement other ones are available in our
repository\footnote{\url{http://git.io/labwiki-plan}}.


\subsection{Publish as a Practical Lab for a Course}

Many recent contributions discussed the benefits of using interactive online
materials and eBooks in teaching Computer Science (CS)
courses~\cite{wright2012,Zhuang_2014,jourjon2010}. However, the creation of
interactive elements (e.g. a practical lab) within an eBook platform currently
requires considerable programming skills as well as familiarity with other
mechanisms, such as student authentication. \labwiki  provides
a simple yet effective solution to generate such an interactive element from an
existing experiment.  Using this solution, a CS lecturer could develop an
experiment in \labwiki, which would involve resources on distributed testbed
facilities and illustrate some aspect of a
course. Once finalized, this experiment could then be ``embedded'' within an
eBook. The students using this eBook would then be able to execute a real experiment trial
through the eBook interface, i.e., a trial actually using real testbed resources.

\begin{figure}
 \centering
 \includegraphics[width=0.4\textwidth]{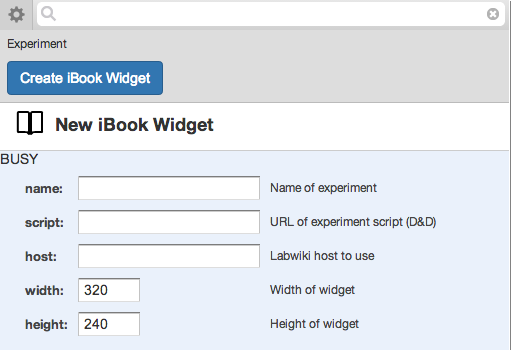}
 \caption{Creating an eBook Widget.}
 \label{fig:widget}
\end{figure}

\descr{Lecturer Side.} Once the lecturer is satisfied with her experiment, she
needs to generate an eBook widget~\cite{langer2012} to act as the interface to
that experiment. Such a widget is a self-contained HTML5/Javascript wrapper,
which can be embedded in an ePub3 or Apple iBook document.  The ``Execute''
panel of \labwiki provides a ``create widget'' action as shown in
Figure~\ref{fig:widget}. The lecturer triggers this action and provides some
configuration parameters for the widget to be generated (e.g., a name, the
display size of the widget in pixel, the set of allocated resources).
The ``create widget'' process then
generates a fully configured widget, which will be downloaded to the lecturer's
machine. She can then include this HTML5/Javascript widget into her eBook using
third-party authoring tools.

\descr{Student Side.}
A student may then download the eBook and open the page with the forementioned widget. Once triggered inside the eBook, the widget will switch to a full-screen web container connected to the remote \labwiki workspace. From then on, the student has access to a subset of the previously mentioned \labwiki features, namely the experiment trial execution (Section~\ref{sec:exp:run}), the result analysis (Section~\ref{sec:result:analysis}), and the publishing (Section~\ref{sec:pub:storing}) features. Thus, she is not allowed to modify the experiment description or the set of resources to use. Once the experiment trial finishes, she may perform any result analysis requested by the lecturer as part of the practical lab, and submit the answers as a \labwiki generated document (Section~\ref{sec:lw:plan}).


\section{Conclusion}
\label{sec:Conclusion}

This paper described a methodology based on well established principles to plan,
prepare and execute experiments.
We proposed and described a family of tools, the \labwiki workspace, to support an experimenter's workflow
based on that methodology. \labwiki enables repeatable experiment-based
research in Computer Networking, Distributed Systems, and to certain extends Computer Science in general.  We
showed how this set of tools leverages large-scale Future Internet initiatives, such as  GENI
and FIRE, and de-facto protocol and model standards, which emerged from these
initiatives. It provides services and mechanisms for each step of an
experiment-based study, while automatically capturing the necessary
information to allow peers to repeat, inspect, validate and modify prior
experiments. Finally, the \labwiki workspace also provides tools
for sharing all generated artifacts (e.g. documents, data, analyses) with the
community. For educators, a seamless mechanism to turn an experiment in an interactive
practical lab for teaching Computer Science is provided.

\section*{Acknowledgments}
NICTA is funded by the Australian Government through the Department of Communications 
and the Australian Research Council through the ICT Centre of Excellence Program.
This material is based in part upon work supported by the GENI (Global Environment for 
Network Innovations) initiative under a National Science Foundation grant.

%
\printbibliography

\end{document}